\documentclass[epj]{svjour}
\usepackage{graphics}
\begin{document}
\title{Phase diagrams of bosonic $AB_{n}$ chains}

\author{G. J. Cruz\inst{1} \and R. Franco\inst{2} \and J. Silva-Valencia\inst{2} 
}                     

\institute{Departamento de Ciencias B\'{a}sicas, Universidad Santo Tomas, Bogot\'{a}, Colombia. \and 
Departamento de F\'{i}sica, Universidad Nacional de Colombia, Bogot\'{a}, Colombia.}
\date{Received: date / Revised version: date}
%
\abstract{
The $AB_{N-1}$ chain is a system that consists of repeating a unit cell with $N$ sites where between the $A$ and $B$ sites there is 
an energy difference of $\lambda$. We considered bosons in these special lattices and took into account the kinetic energy, the local two-body 
interaction, and the inhomogenous local energy in the Hamiltonian. We found the charge density wave (CDW) and superfluid and Mott insulator phases, and 
constructed the phase diagram for $N=2$ and $3$ at the thermodynamic limit. The system exhibited insulator phases for densities $\rho=\alpha/N$,
with $\alpha$ being an integer. We obtained that superfluid regions separate the insulator phases for densities larger than one. 
For any $N$ value, we found that for integer densities $\rho$, the system exhibits $\rho +1$ insulator phases, 
a Mott insulator phase, and $\rho$ CDW phases. For non-integer densities larger than one, several CDW phases appear.
} 
\maketitle
\section{Introduction}
\label{intro}
The study of the ground state of bosonic systems is an interesting area within  the study of cold atoms due to the capability of emulating them in 
optical lattices, with the advantage of having absolute control over the parameters (kinetic energy, interactions, and density) and access to various 
dimensions~\cite{bloch1,bloch2}.\par
Experimental progress on this topic has allowed the prediction of the superfluid-Mott insulator transition in ultracold bosonic atoms~\cite{jack}
and the observation of quantum phase transitions~\cite{gre} in an optical lattice, which has opened the way for potential studies that have revealed the 
fundamental physics of these systems.\par 
Recently, the implementation of optical lattices has motivated the exploration of bosonic systems in superlattices whose arrangement is characterized 
by a periodic potential~\cite{Roati08,Atala13}, for instance an energy difference of $\lambda$ in the unit cell between sites called $A$ and $N-1$ 
$B$ in the unit cell (notation $AB_{N-1}$). Bidimensional systems with $AB$ configurations have been created confining 
$^{87}$Rb~\cite{Liberto} and $^{40}$K~\cite{messer} atoms in optical lattices. This experimental progress allows us to believe that experimental 
study of the one-dimensional $AB_{N-1}$ chains will be carried out in the coming years.\par
In 2004, Buonsante and Vezzani~\cite{buon}  elaborated an analytical description of the physics of cold bosonic atoms trapped in superlattices. 
Within the study of the finite-temperature phase diagram, they included the results for zero temperature, and insulator domains for fractional 
densities were demonstrated. The phase diagrams due to the interplay between the on-site repulsive interaction, superlattice potential strength, 
and filling were found by Rousseau et al. ~\cite{Rousseau}.\par
\begin{figure}[t]
\resizebox{0.45\textwidth}{!}{%
  \includegraphics{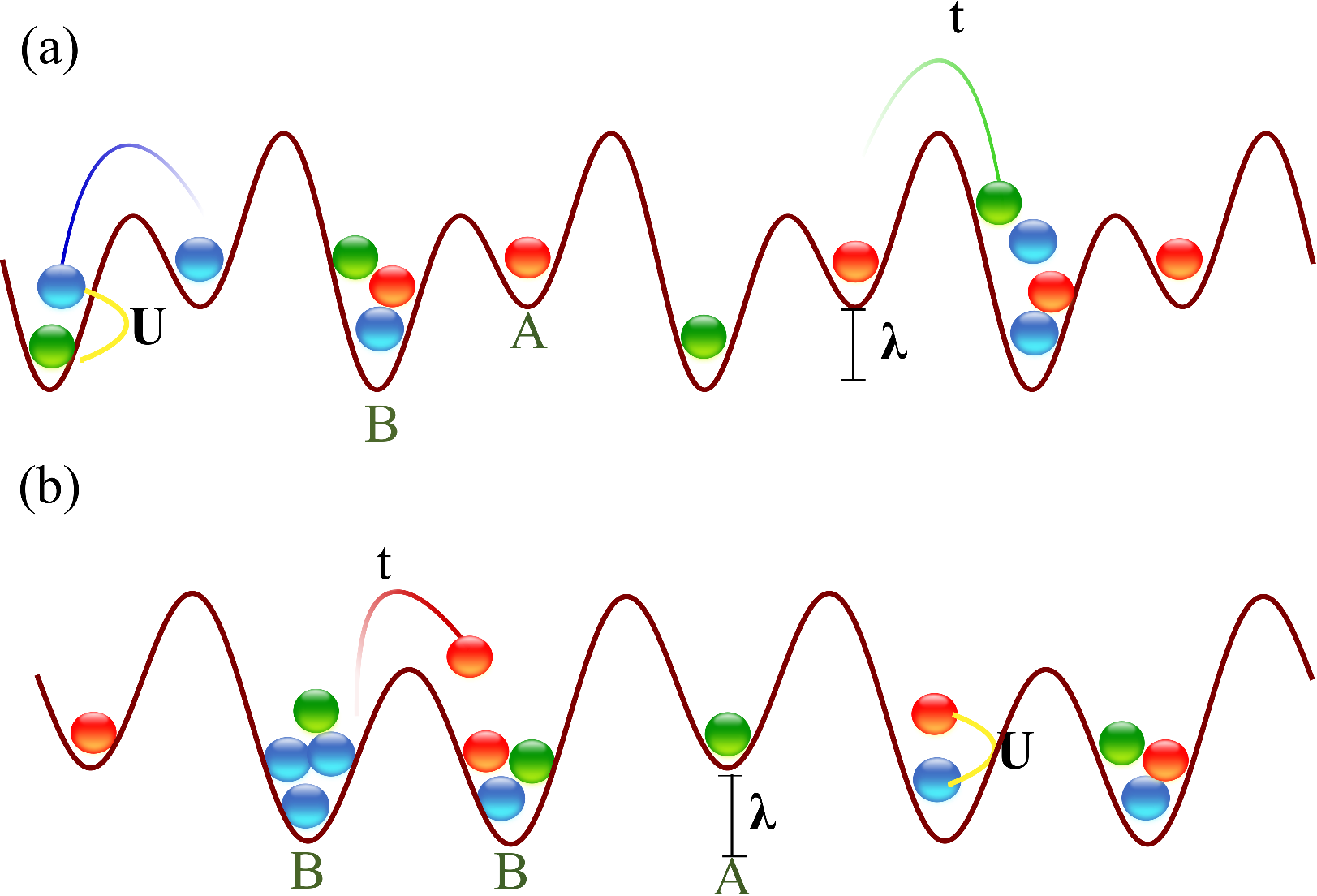}
}
\caption{(color online). Schematic representation of the superlattices considered in this paper.  The chain $AB$ (a) is formed by 
 the superposition of two waves, where the first has double the frequency of the other, while the chain $AB_{2}$ in
 (b) is formed by the superposition of two waves where the first has triple the frequency of the other. We consider that the difference of 
 the hopping parameter between neighboring states is very small.}
\label{super}       
\end{figure}
Later, Dhar and colleagues~\cite{dhar1,dhar2,dhar3} studied the quantum phases that emerge when the $AB$ chain is considered (see Fig. \ref{super}(a)). 
Such studies proved that a new insulator phase results at density  $\rho=1/2$, and the Mott insulator phase corresponding to $\rho=1$ undergoes a 
phase transition to another insulator when the parameter $\lambda$ is near the strength of the interaction between the particles ($U$), leading to a 
different boson arrangement.\par
Although the results of these new phases indicate the consequences produced by the superlattice on a boson chain, 
we believe that there are new aspects to be discovered. For instance, we can ask: is it possible that the insulator phases at $\rho>1$ evolve 
to other insulator phases passing briefly through a superfluid when $\lambda$ increases? What happens in the phase diagram if we increase the 
number of sites in the unit cell?
To answer these questions, we used the density matrix renormalization group method~\cite{white} and the von Neumann entropy~\cite{sou11} 
to study the quantum phases of the model. We found that for fractional densities lower than one there is a unique CDW phase for any value of 
$\lambda$. For densities larger than one, we obtained a finite number of insulating phases when $\lambda$ increases, two CDW phases for fractional 
densities, and $\rho +1$ for integer ones. The von Neumann entropy shows that the insulator phases are separated by superfluid regions.\par
This paper is organized in the following way: In section 2, we explain the Hamiltonian for $AB_{N-1}$ chains. The thermodynamic limit phase diagrams and the 
von Neumann entropy results for $AB$ and $AB_2$ chains are shown in sections 3 and 4. Finally, the conclusions are presented in section 5.\par
\begin{figure}[t]
\resizebox{0.45\textwidth}{!}{%
  \includegraphics{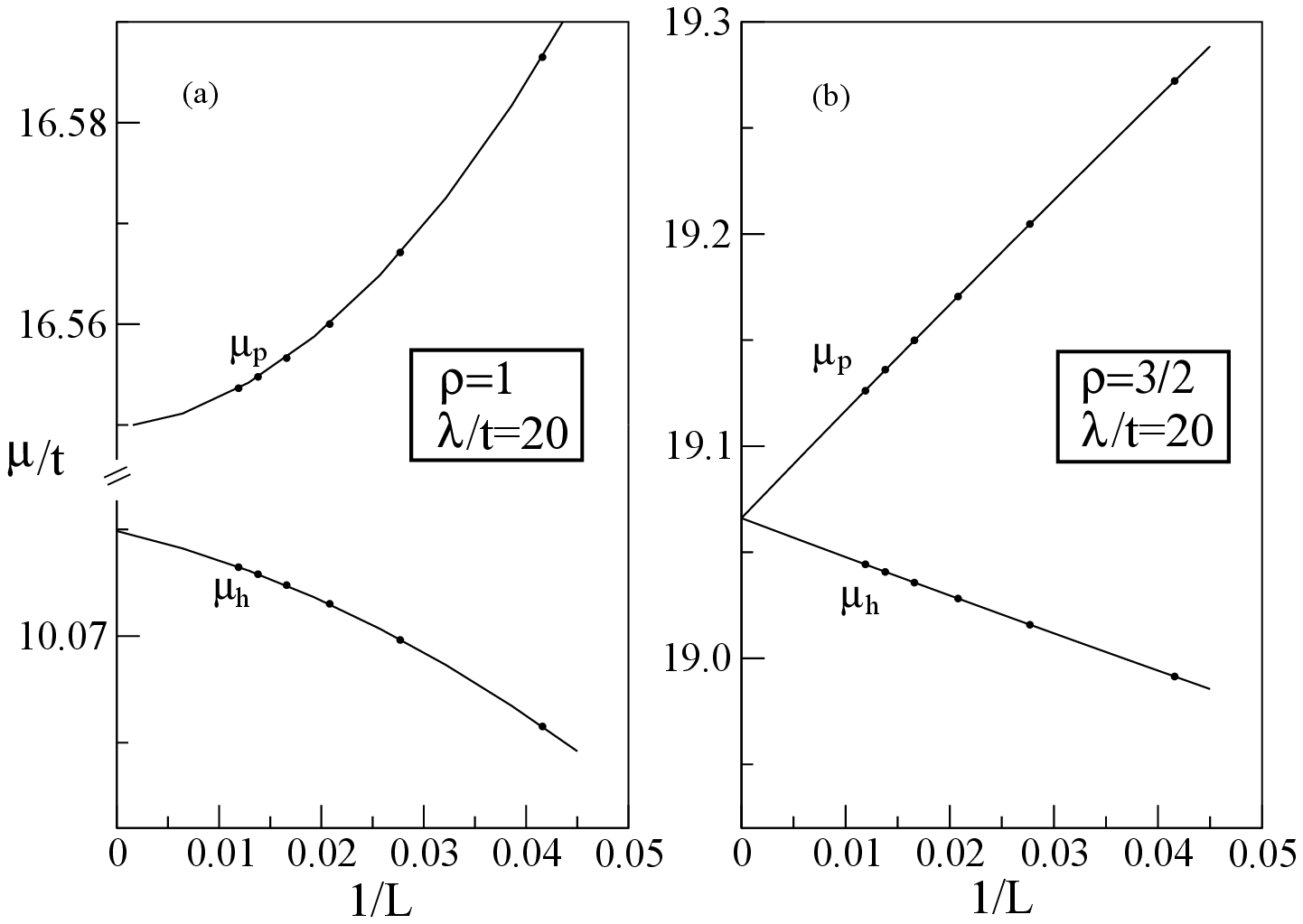}
}
\caption{Graphs of chemical potential versus $1/L$ for $\lambda/t=20$ in a) $\rho=1$ and b) $\rho=3/2$. The points are DMRG results, and the lines 
  represent the regression to the thermodynamic limit.}
\label{limter}       
\end{figure}
\section{Model}
\label{sec:sec2}

The ground state of ultracold bosonic atoms in homogeneous systems is described by the Bose-Hubbard model. When a system of bosons in an 
inhomogeneous lattice is considered, the Hamiltonian is given by:

\begin{eqnarray}
 H&=&-t \sum _{<i,j>}(\hat{a}_{i}^{\dagger}\hat{a}_{j}+ H.c.)+\frac{U}{2}\sum_{i}\hat{n}_{i}(\hat{n}_{i}-1)  \nonumber \\  
   & & +\sum_{i}\lambda_{i}\hat{n}_{i} - \mu\sum_{i}\hat{n}_{i}.
\label{H1}
\end{eqnarray}

\noindent $t$ being the hopping parameter, $<i,j>$ denotes a pair of nearest-neighbor sites $i$, and $j$, $\mu$ is the chemical potential, 
$a_{i}^{\dagger}$ $(a_{i})$ creates (annihilates) a boson at site $i$. $U$ represents the local interaction in the second term of the Hamiltonian, 
where $\hat{n}_{i}=a_{i}^{\dagger}a_{i}$ is the  number operator, and $\lambda_{i}$ denotes the shift in the energy levels of the sites in each 
unit cell. We set our energy scale taking $t=1$ and the interaction parameter $U/t=10$.\par
\begin{figure}[t]
\resizebox{0.45\textwidth}{!}{%
  \includegraphics{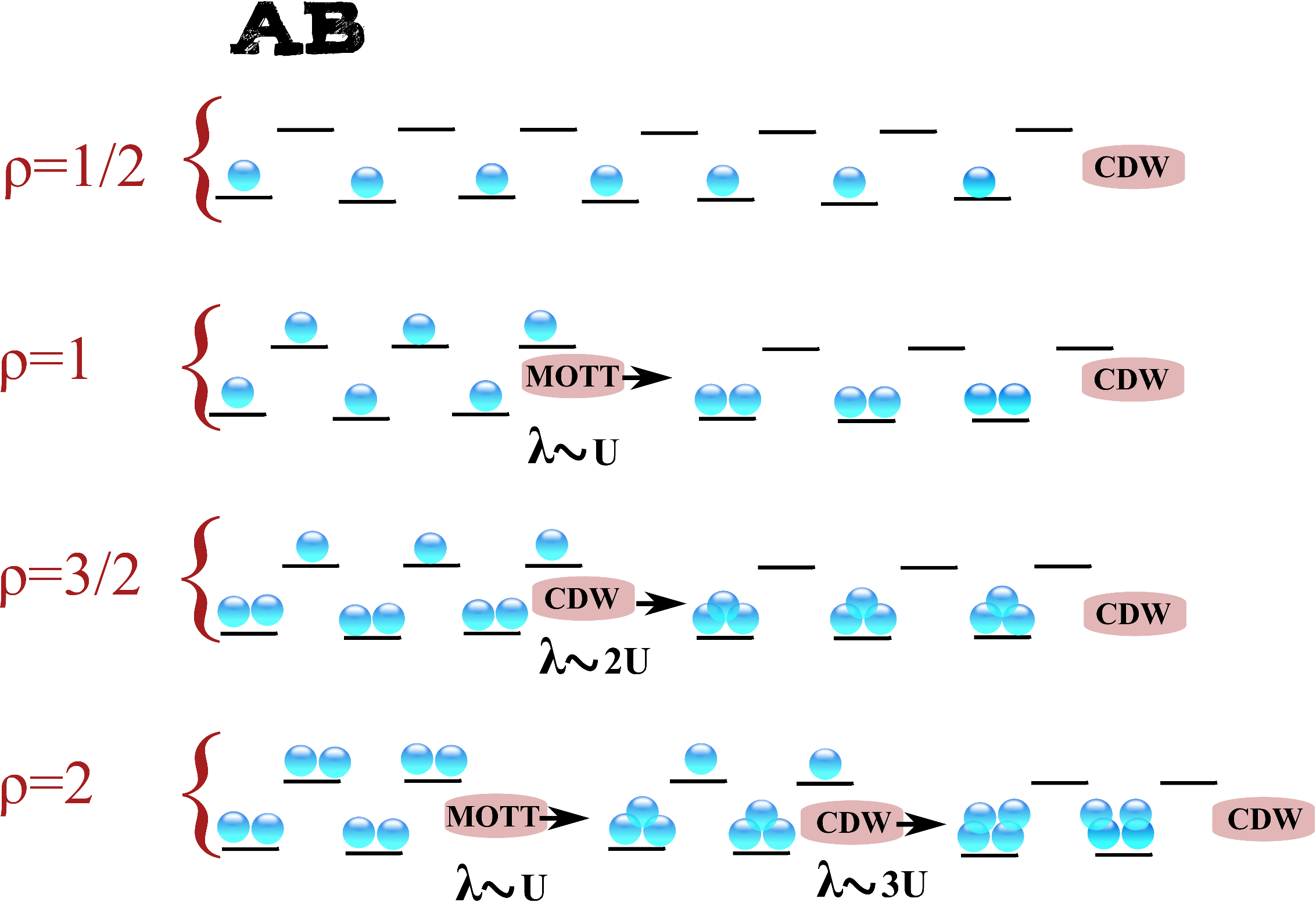}
}
\caption{(color online). Phases and transitions in the $AB$ chain.}
\label{AB}       
\end{figure}
Particle chains in a superlattice are denoted as $AB_{N-1}$, where the site $A$ has a difference of energy of $\lambda$ from the $N-1$ sites called 
$B$ per unit cell.
In the present paper, we consider two types of superlattices. The first superlattice has a potential with periodicity $2$ ($AB$ chains), meaning two sites 
per unit cell with a potential difference $\lambda$, schematically represented in Fig. \ref{super}(a). The second one has a periodicity of $3$
and is shown in Fig. \ref{super}(b) ($AB_{2}$ chains).\par
As is well known, in a homogeneous environment the one-dimensional bosonic system exhibits a phase transition between a Mott insulator phase, 
characterized by integer filling, and the superfluid phase, which is compressible~\cite{kun1,kun2,eji}. In a superlattice type $AB$,
the behavior of the system in the ground state exhibits additional phases for half-filling and density equal to one,
where insulating phases due to the superlattice potential are exhibited~\cite{dhar1}.\par 
The current investigation involved calculating the energies $E(N,L)$ for lattices with different lengths $L$ and $N$, $N+1$ and $N-1$ particles 
such that we obtain the chemical potential for the increase $(\mu_{p})$ and decrease $(\mu_{h})$  the number of particles to one, where the 
general expressions are: 
\begin{figure}[t]
\resizebox{0.45\textwidth}{!}{%
  \includegraphics{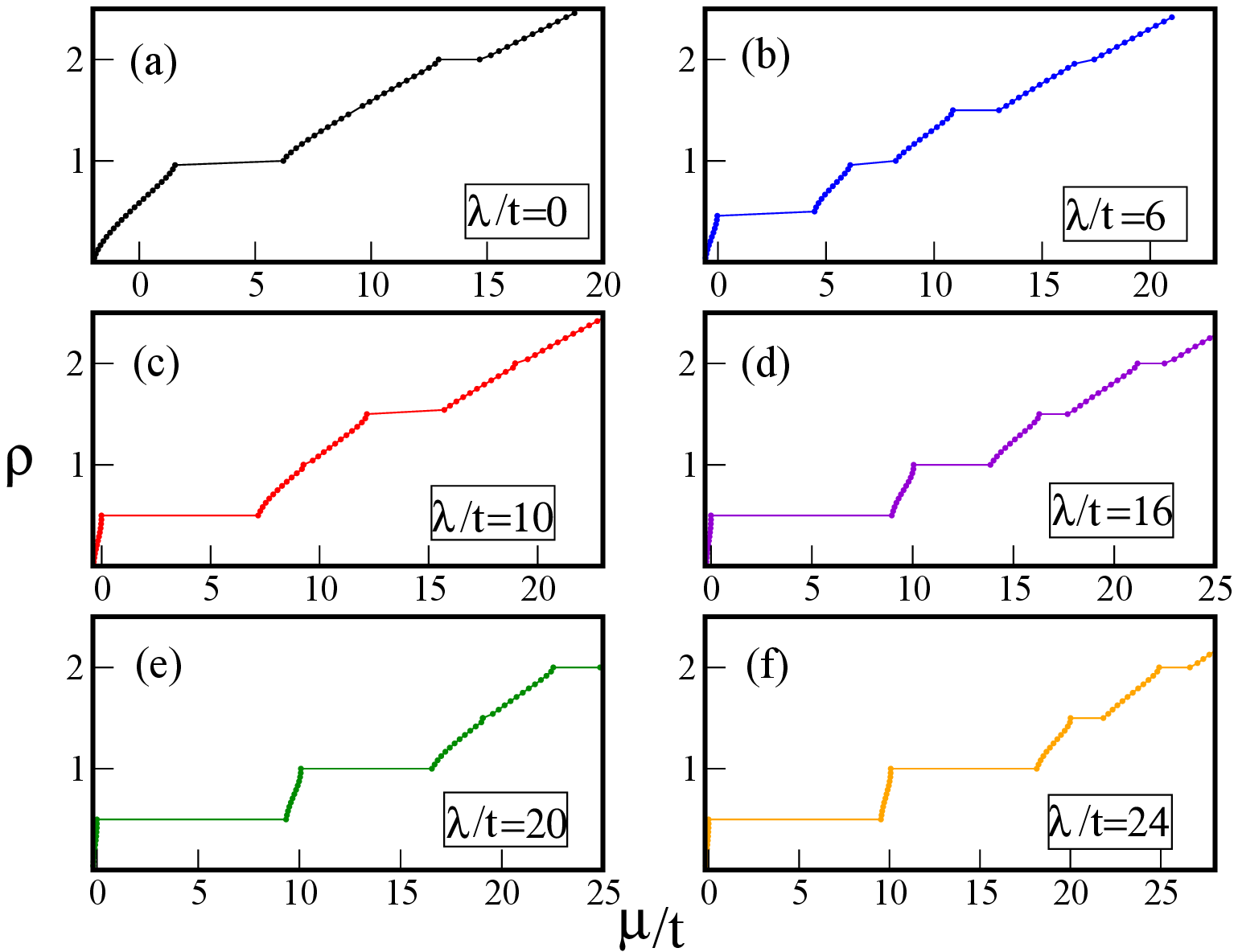}
}
\caption{(color online). Density profile $\rho$ versus chemical potential $\mu$  for bosons in superlattice type AB and various values 
  of $\lambda$, with $U/t=10$.}
\label{rhomiu2}       
\end{figure}
\begin{equation}
\mu_{p}=E(N+1,L)-E(N,L)
\end{equation}
and
\begin{equation}
\mu_{h}=E(N,L)-E(N-1,L).
\end{equation}
To obtain the energies, we used the density matrix renormalization group method for lattices from 24 to 84 sites, obtaining an error of around 
$10^{-13}$ with open boundary conditions. Then we extrapolated our results at the thermodynamic limit and repeated the process for a wide range of values 
of the parameter $\lambda$. 
 For example, all insulator phases are characterized by a gap. The analysis of this quantity at the limit $L\rightarrow\infty$ gives us information 
 about the phase in which the system exists.
The gap for any boson system is given by:
\begin{equation}
\Delta=\mu_{p}-\mu_{h}.
\end{equation}

\noindent If $\Delta\neq0$, the system is in an insulator phase; if it is not, the system is in a superfluid phase. This is shown in Fig. \ref{limter} for 
the $AB$ chain for two cases: in Fig. \ref{limter}(a), we consider a density $\rho=1$ and $\lambda/t=20$ and we observe that the 
system has a large gap at the thermodynamic limit, which indicates that the system is in a CDW phase, and specifically the charge 
distribution generated by the superlattice structure is $\{2,0,2,0,2,0...\}$, as reported before (see Fig. \ref{AB}). If we maintain the $\lambda$ 
parameter the same and change the global density to $\rho=3/2$, we expect that the system will be in a CDW state due to the superlattice, according to 
the results of Buonsante and Vezzani\cite{buon}, with a density profile $\{2,1,2,1,2,1...\}$(see Fig. \ref{AB}). However, in Fig. \ref{limter}(b), we 
observe a gap equal to zero at the thermodynamic limit, which indicates that the system is in a superfluid state. In conclusion, for $\lambda/t=2U$, the 
ground state is CDW for $\rho=1$ and is superfluid for $\rho=3/2$.\par
\begin{figure}[t]
\resizebox{0.45\textwidth}{!}{%
  \includegraphics{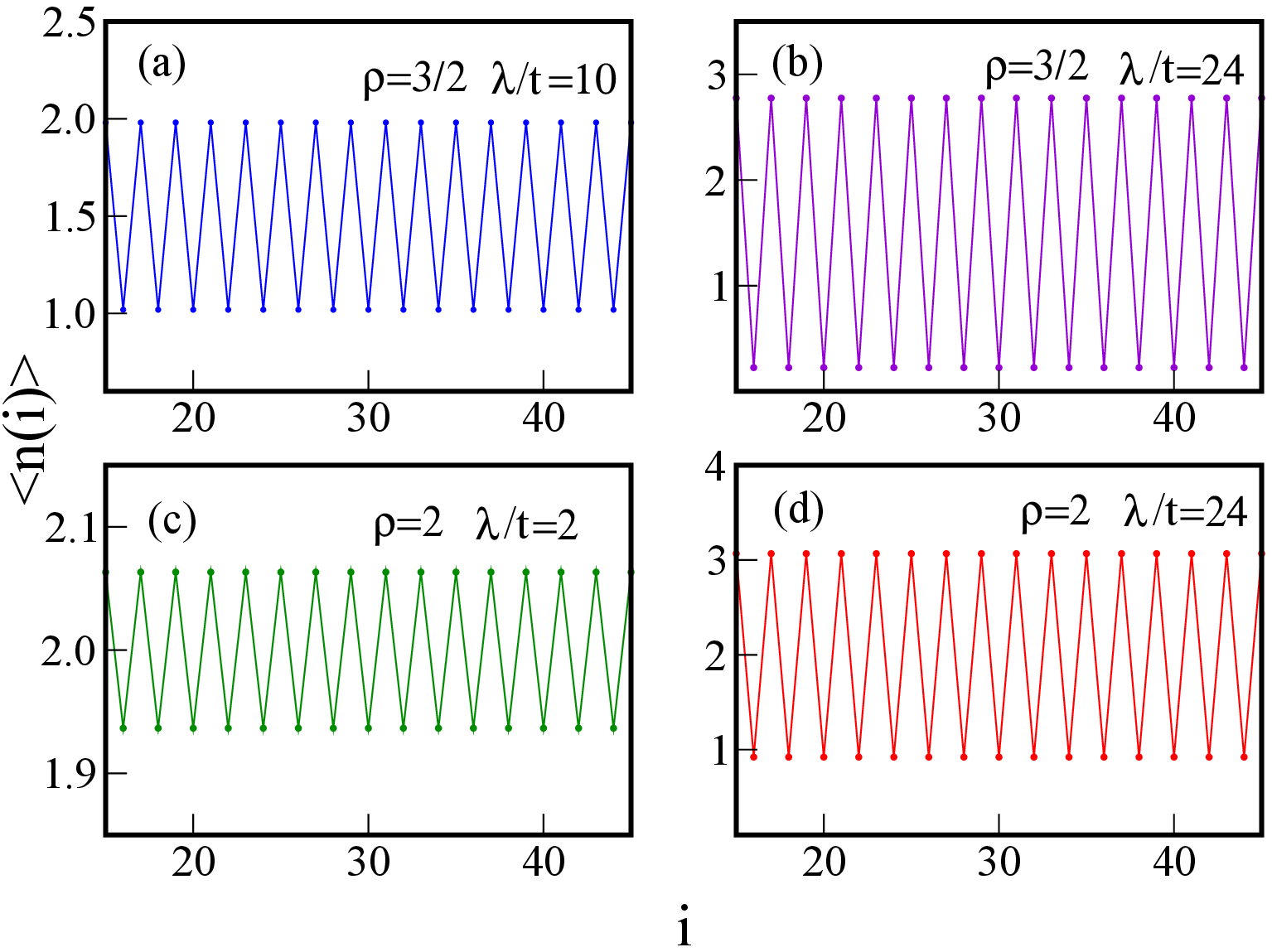}
}
\caption{(color online). On-site number density plotted against lattice site index with $L=50$ sites at density $\rho=3/2$ and $\rho=2$. 
Open boundary conditions were used. The lines are visual guides.}
\label{occup2}       
\end{figure}
\section{$AB$ phase diagram}
\label{sec:sec4}

Dhar \textit{et al.} \cite{dhar1} have shown that considering  bosons in a superlattice with periodicity equal to two, a CDW phase with one particle 
per unit cell appears for $\rho=1/2$ (see Fig. \ref{AB}), which remains for any value of $\lambda$, while for $\rho=1$, they obtained a Mott insulator 
phase and a CDW phase, because when the hopping and the local interaction parameters are fixed, for values of $\lambda<U$ the ground state will be a Mott 
insulator with one particle per site. However, for values of $\lambda \approx U$, the local repulsion can be compensated by the superlattice term, and the 
quantum fluctuations increase, delocalizing the particles, which leads to a superfluid state. As $\lambda$ is increased further, the bosons prefer to be 
in the wells, and the ground state is a CDW (see Fig. \ref{AB}). They showed that the superfluid region between the Mott insulator and the CDW phases is 
in the range $\{9.6,10.0\}$. In this section, we extend these results by considering densities larger than $\rho=1$ in the $AB$ chain.\par
In Fig. \ref{rhomiu2}, we show the density $\rho$ versus the chemical potential $\mu/t$ for some specific values of $\lambda/t$, 
considering $U/t=10$, because the previous phase diagram was constructed for this value. The case of a homogeneous lattice, with $\lambda/t=0$, 
is displayed in Fig. \ref{rhomiu2}(a). In this figure, it is possible to clearly observe  plateaus at integer densities of $\rho=1$ and $\rho=2$. 
The width of the plateaus indicates the size of the energy gap at the thermodynamic limit for each density. Between these values of $\mu/t$, 
a Mott insulator phase is found. For different values of $\mu/t$, the system is compressible and is in a superfluid phase.\par 
When an inhomogeneous lattice is considered, for example with $\lambda=6$, shown in Fig. \ref{rhomiu2}(b), the number of plateaus increases. It is 
crucial to emphasize that the plateaus for $\rho=1$ and $\rho=2$ are smaller than for $\lambda=0$, where the plateau for $\rho=1$ is larger than 
the plateau for $\rho=2$. Also, the other plateaus for half-integer $\rho$ are shown as predicted. These plateaus are caused by the superlattice 
structure type that characterizes the boson chain, while the plateaus for integer densities are caused by the interaction between the particles of 
each site. It is important to highlight the fact that between each plateau at a specific value of $\lambda/t$, we always obtain a superfluid 
region.\par 
In Fig.  \ref{rhomiu2}(c), for $\lambda/t=10$, plateaus for densities $\rho=1/2$ and $\rho=3/2$ are observed, while for $\rho=1$ and $\rho=2$ the 
width of the plateaus is very small, and around $\lambda \approx U$ we found that the Mott insulator phase disappears for all integer densities and 
a superfluid region anticipates a new insulator phase. On the basis of Fig. \ref{AB} for $\rho=2$, we can explain why the Mott insulator phase disappears. A 
Mott insulator phase means $\rho$ particles per site, which implies an energy $U\rho(\rho-1)+\rho\lambda$ for two sites. For $\lambda \approx U$, this 
energy can be compensated by a particle jump into wells, the energy for two sites is  $U(\rho^2-\rho+1)+(\rho-1)\lambda$, and the fluctuations lead to a 
superfluid state for any density $\rho$. Note that the hopping term is small ($U/t=10$). This contributes to the fluctuations, but is not 
the most important factor. This fact should be taken into consideration throughout this paper.\par
\begin{figure}[t]
\resizebox{0.45\textwidth}{!}{%
  \includegraphics{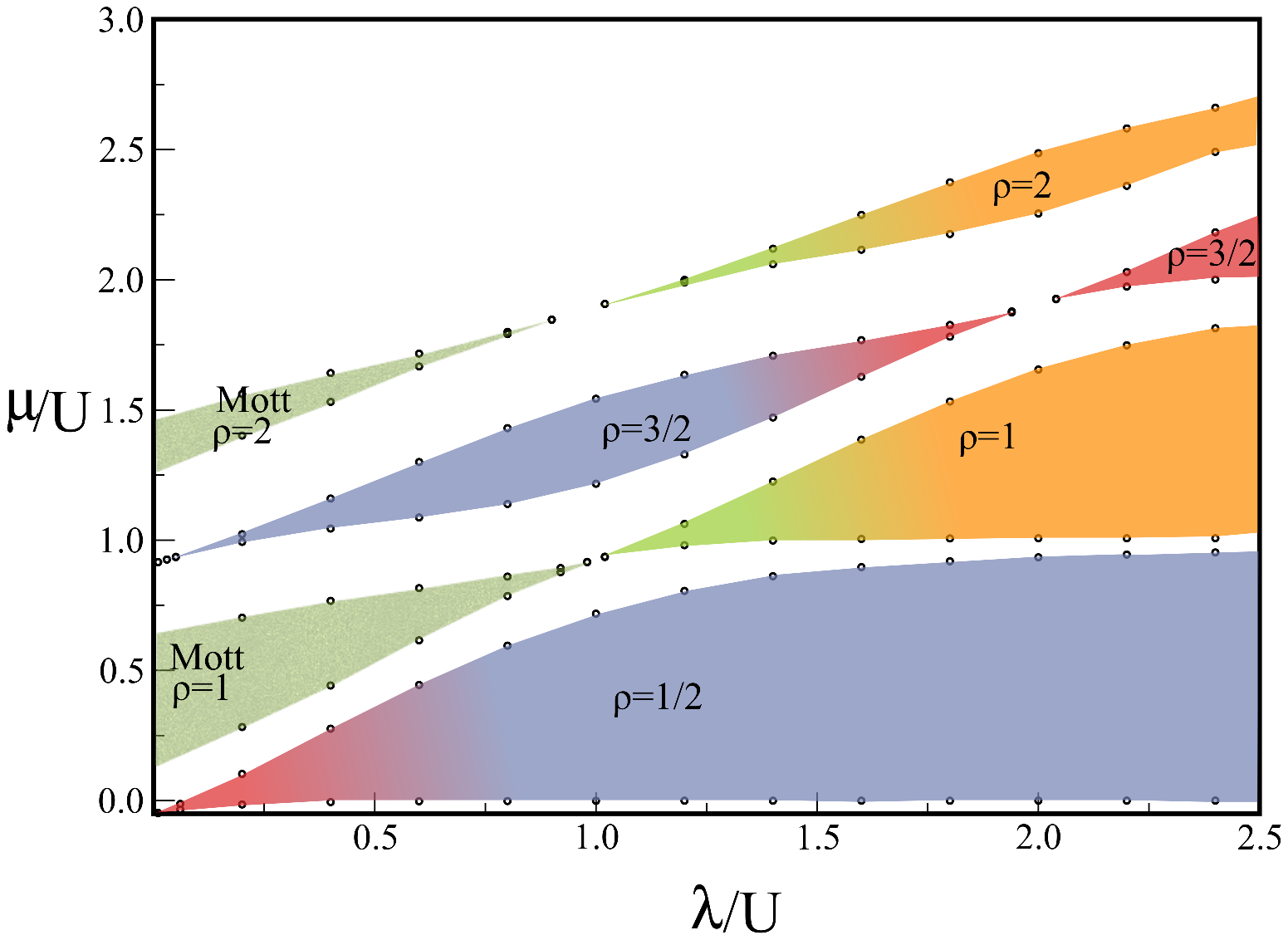}
}
\caption{(color online). Phase diagram of one-dimensional bosonic system for the $AB$ chain. The points (open circles) represent the boundaries and 
  are DMRG results.}
\label{n2}       
\end{figure}
For $\lambda/t=16$, the plateaus reappear, and we get insulator phases at all commensurate densities $\rho=1/2, 1, 3/2$ and $\rho=2$. Also, we can 
observe that the plateau for $\rho=1/2$ always increases as $\lambda/t$ increases, which is shown in Fig. \ref{rhomiu2}(d). In Fig.  \ref{rhomiu2}(e), 
for $\lambda/t=20$ the plateau at $\rho=3/2$ is not present, and the insulator phase due to the superlattice disappears around $\lambda \approx 2U$. 
From Fig. \ref{AB}, we see that the CDW phase for this density consists of two bosons in the wells and one in the barriers. Note that if the barrier 
particle jumps to the well, the energy for two sites $3U$ compensates for the energy of the CDW arrangement $U+\lambda$, when $\lambda \approx 2U$, which 
leads the system to a superfluid state.\par
Finally, for $\lambda/t=24$, it is possible to observe that again we have plateaus for all the densities, and a new 
insulator phase for $\rho=3/2$ is obtained, as is shown in Fig. \ref{rhomiu2}(f).\par
The above discussion allows us to conclude that for all densities $\rho\geq1$, a new CDW insulator phase occurs for larger values of $\lambda/t$. For 
integer densities, this CDW phase is achieved by first passing through a Mott insulator phase and a superfluid one, while for $\rho=3/2$, we 
obtained two CDW phases separated by a superfluid phase.\par 
\begin{figure}[t]
\resizebox{0.45\textwidth}{!}{%
  \includegraphics{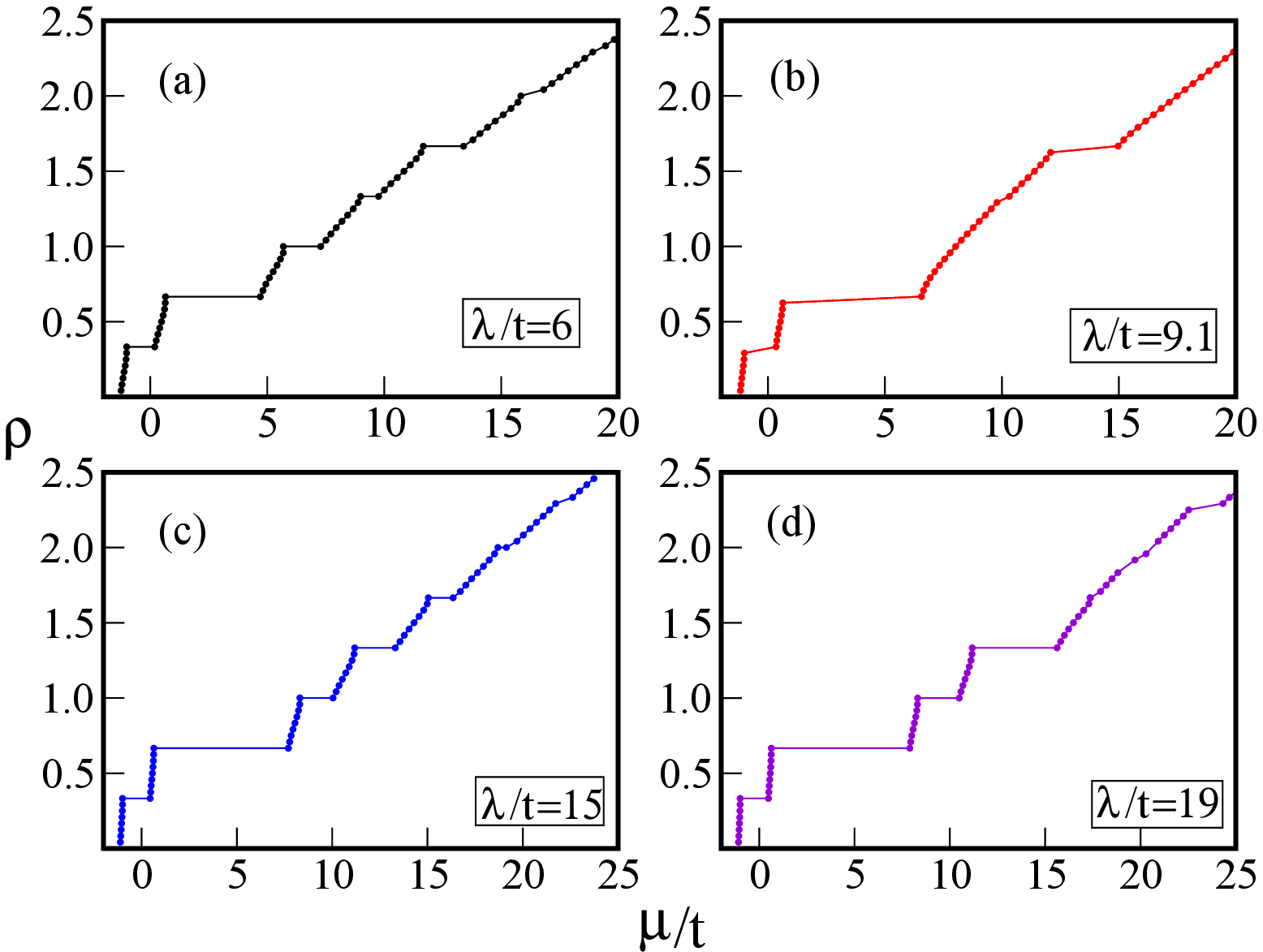}
}
\caption{(color online). Density versus the chemical potential for some values of $\lambda/t$, for a system of bosons in a superlattice type $AB_{2}$.}
\label{rhomiu3}       
\end{figure}
Each of the above insulator phases that have been mentioned is characterized by a particular charge distribution in the system. In Fig. \ref{occup2}, 
the lattice density profile is shown for densities $\rho=3/2$ and $\rho=2$, with some parameters taken before and after the transition region. In 
Fig. \ref{occup2}(a), for $\rho=3/2$ and $\lambda/t=10$, the distribution of two bosons at one site and one boson at the other site is observed, 
while for $\lambda/t=24$, in Fig. \ref{occup2}(b), the charge distribution is close to $\{3,0,3,0, 3, 0...\}$ . Clearly, the system experiences a 
change of state caused by the spatial structure of the lattice. We emphasize that this possibility was not taken into account by Dhar \textit{et al.}\par
In Fig. \ref{occup2}(c) for $\lambda/t=2$, a Mott insulator phase with two particles at each lattice site is shown, but as $\lambda/t$ increases 
to $\lambda/t=24$ in Fig. \ref{occup2}(d), the charge distribution in the system indicates that the system is organized as $\{3,1,3,1,3,1...\}$.\par  
As was reported, the insulator phase at density $\rho=1/2$ for $\lambda/t \neq 0$ is always present, and for $\rho=1$, a phase transition 
is located near $\lambda/t\approx U$ (see Fig. \ref{n2}). Also, this figure shows that for all commensurate densities with $\rho \geq 1$, the system 
exhibits a superfluid region between two insulator regions, and the parameter $\lambda$ can generate a quantum transition between these states due 
to the superlattice structure. Within the values of $\lambda/t$ considered, we see that there are two insulator phases for densities $\rho\geq1$, 
and the gap of the CDW phase for the densities $\rho=1/2$ and $\rho=1$ saturates, because for these densities, the barriers are unoccupied (see 
Fig. \ref{AB}), and this implies that no state change may occur.\par

\section{$AB_{2}$ phase diagram}
\label{sec:sec5}
Now the case of bosons in a superlattice type $AB_{2}$ will be studied. In this superlattice, each unit cell is formed by three sites. 
Two $B$ sites differ by a potential of $\lambda$ with respect to $A$ (see Fig. \ref{super}(b)).  Note that real systems with this structure in 
one or three dimensions have been created and studied~\cite{yama,redl,torio}.\par
Again, we consider a local repulsion interaction between the bosons of $U/t=10$, and the chemical potential is determined at the thermodynamic limit.\par
\begin{figure}[t]
\resizebox{0.45\textwidth}{!}{%
  \includegraphics{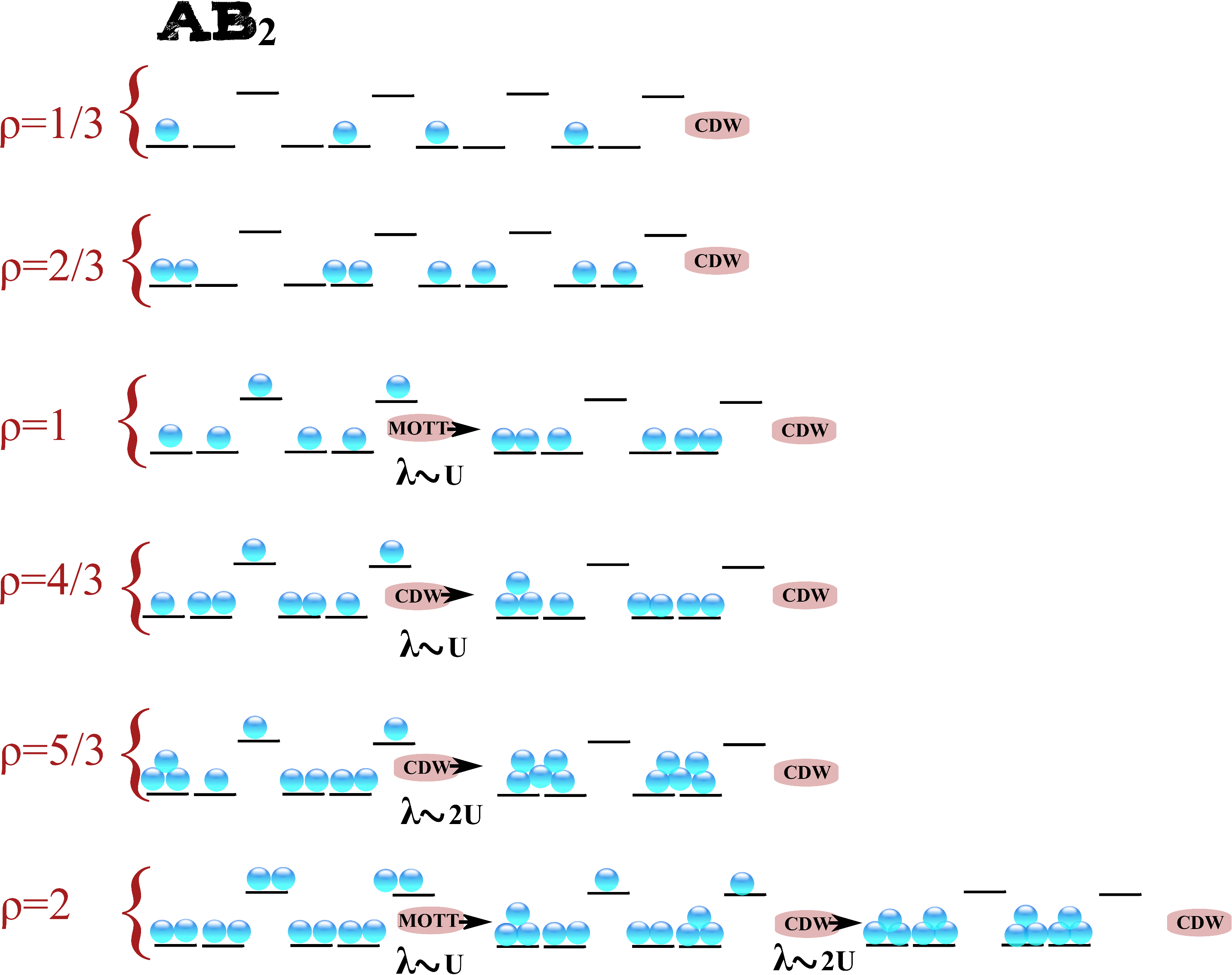}
}
\caption{(color online). Phases and transitions in the $AB_2$ chain.}
\label{AB2}       
\end{figure}

The evolution of the plot of density versus chemical potential as a function of the superlattice parameter is shown in Fig. \ref{rhomiu3}. For 
$\lambda/t=6$ (Fig. \ref{rhomiu3}(a)), the number  of plateaus increases, in contrast with the superlattice type $AB$. The first two plateaus are 
related to the possibility of having one or two bosons in the wells of each unit cell (see Fig. \ref{AB2}), which means CDW phases with densities 
$\rho=1/3$ and $\rho=2/3$. As in the $AB$ case, we expected that these plateaus would increase with $\lambda$ until they saturate. Also, we obtained other 
plateaus at commensurate densities of multiples of $1/3$, as was predicted.\par
Fig. \ref{rhomiu3}(b) shows the results for $\lambda/t=9.1$, 
where the plateaus at $\rho=1$, $\rho=4/3$ and $\rho=2$ disappear, while the size of the gaps for the other commensurate densities remains finite. The 
reappearance of these plateaus happens for larger values of $\lambda$, as can be seen in Fig. \ref{rhomiu3}(c). This means that there is a superfluid 
region separating two insulator regions for $\lambda$ around $U$ for each density. For integer densities, the explanation is similar to the previous 
case ($AB$); fixing the density $\rho$ (each site in the well has $\rho$ bosons) and a barrier strength  $\lambda$, the energy of two sites (the 
barrier and one of the well) is $U\rho(\rho-1)+\rho\lambda$ (see Fig. \ref{AB2}). If one boson jumps into the well,  previous energy can be compensated when  
$\lambda \approx U$. Now the energy of the two sites is $U(\rho^2-\rho+1)+(\rho-1)\lambda$. We can conclude that for any $AB_{N}$ chain, there is always a 
superfluid region around $U$ separating a Mott insulator phase and a CDW one for any density $\rho$.\par 
Finally, in Fig. \ref{rhomiu3}(d), with $\lambda/t=19$, the calculations show that the gaps for $\rho=5/3$ and $\rho=2$ are very small, which 
indicates that for these densities a superfluid region will appear around $2U$. Note that in the $AB$ chain, a superfluid region around $2U$
appears for the commensurate density $\rho=3/2$, but this does not occur for the integer density $\rho=2$ (see Fig. \ref{n2}). For $\lambda=2U$ and $\rho=2$, the 
charge distribution of the ground state is three particles in the well and one in the barrier (see Fig. \ref{AB}), so the quantum transition 
happens when the particle in the barrier jumps to the well. The above happens for $\lambda \approx 3U$. Before the transition, the energy for two sites is 
$3U+\lambda$, which can be compensated with the energy $6U$ once the particle has jumped into the well. The location of this transition marks one of 
our new findings due to the increase of the number of sites of the wells.\par
\begin{figure}[t]
\resizebox{0.45\textwidth}{!}{%
  \includegraphics{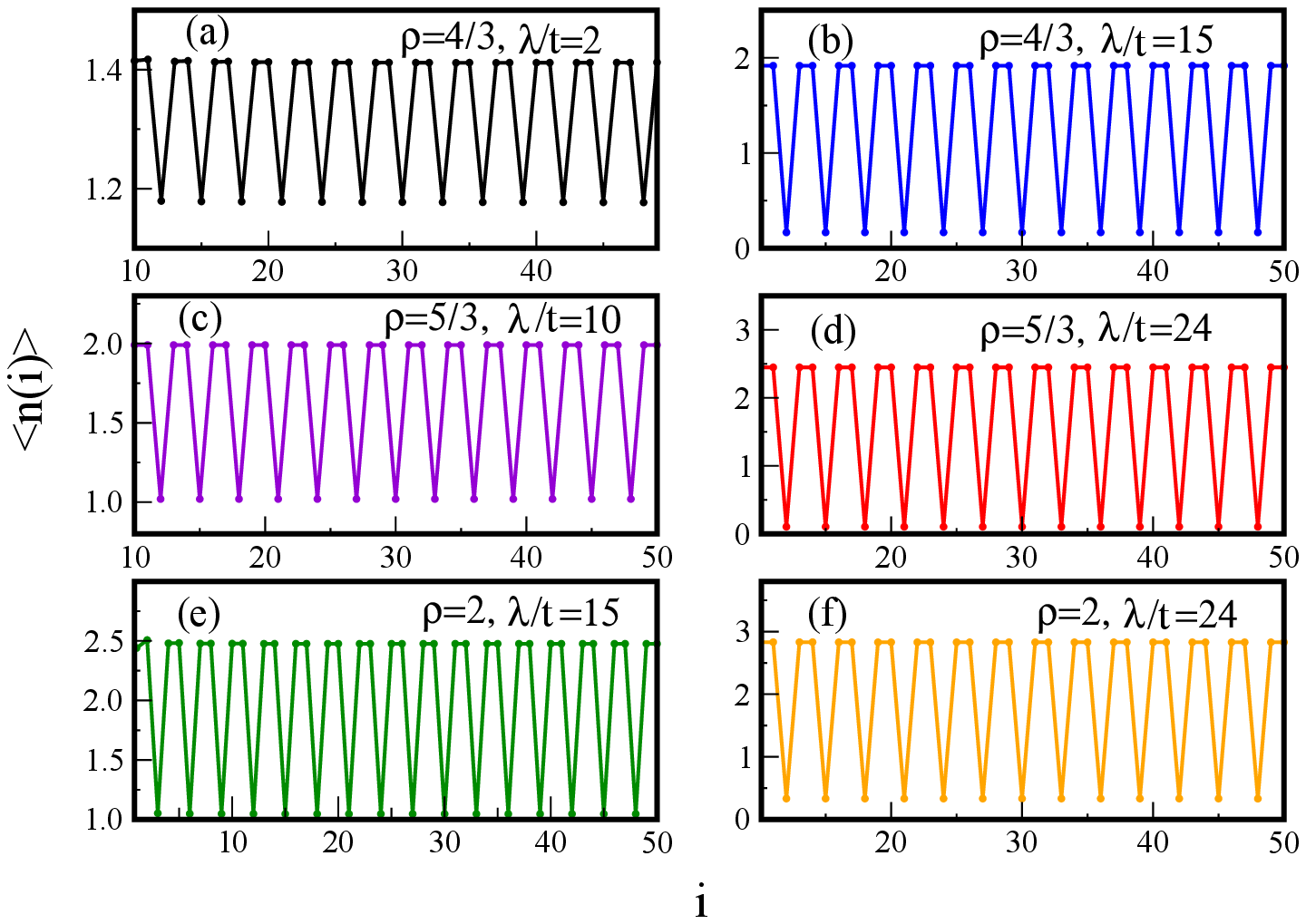}
}
\caption{(color online). On-site number density plotted against lattice site index at densities $\rho=4/3,5/3$ and $\rho=2$ and values smaller 
  or larger than the critical point. Open boundary conditions were used. The lines are visual guides.}
\label{corr3}       
\end{figure}
\begin{figure}[t]
\resizebox{0.45\textwidth}{!}{%
  \includegraphics{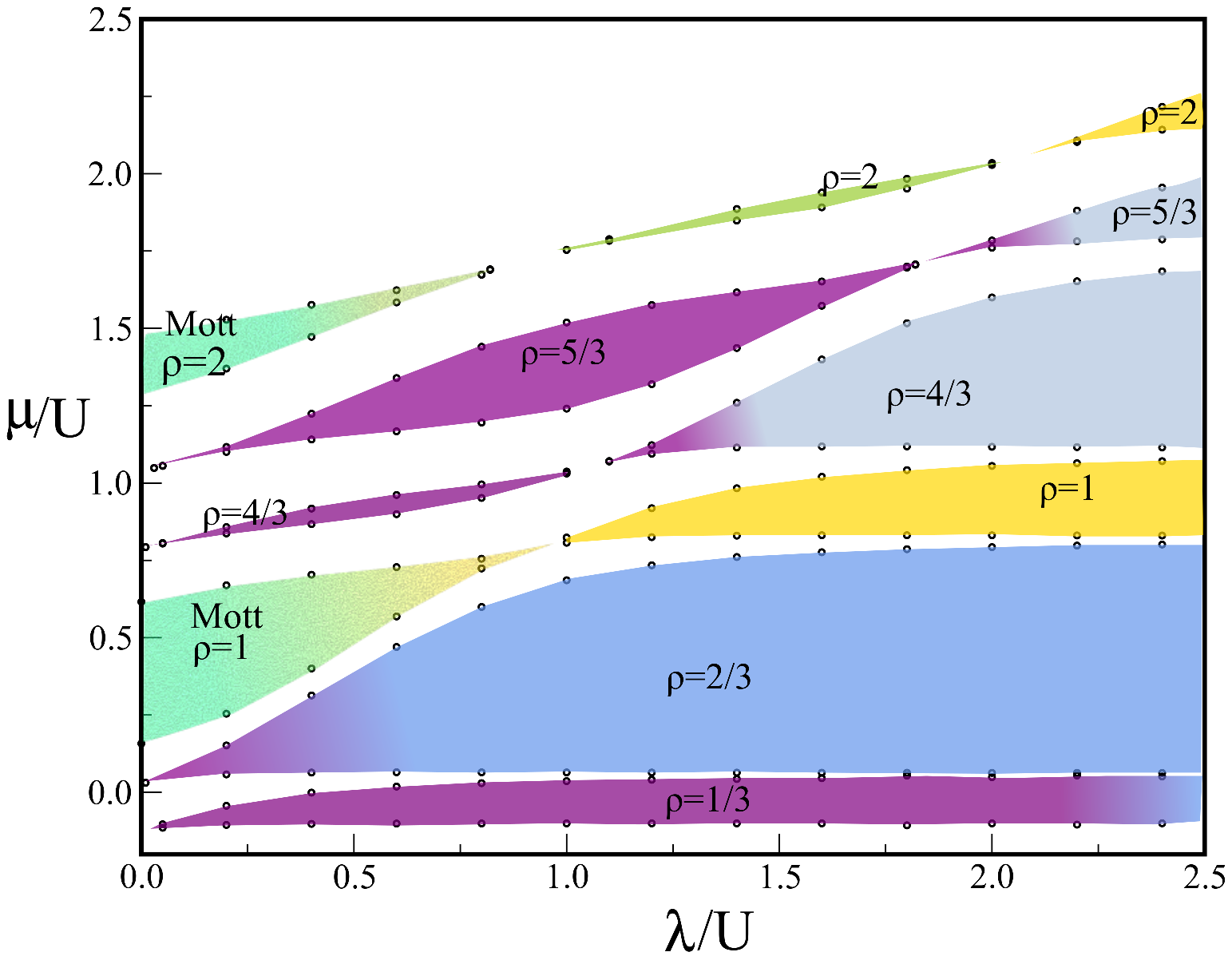}
}
\caption{(color online). Phase diagram for bosons in a superlattice type $AB_{2}$ at the thermodynamic limit. The points (open circles) 
  represent the boundaries in the density matrix renormalization group method.}
\label{diag3}       
\end{figure}

In order to distinguish between insulators, we calculate the arrangement of the particles in the superlattice according to the increase of the 
parameter $\lambda/t$; i. e. the density profiles for values of $\lambda/t$ lower and higher than the transition region are shown in Fig. \ref{corr3}, 
where the densities $\rho=4/3$, $\rho=5/3$ and $\rho=2$ are considered. For density $\rho=4/3$ in Fig. \ref{corr3}(a), with $\lambda/t=2$, the 
superlattice induces fluctuations of charge such that there is one particle in the barrier and three in the well (see Fig. \ref{AB2}), while 
Fig. \ref{corr3}(b), with $\lambda/t=15$, shows the particles organized such that the occupancy is  $\{2,2,0,2,2,0, ...\}$. The quantum phase 
transition takes place around $U$, which can be explained by taking into account that the smallest number of particles in a site within the well is one. The 
energy of the two sites, the barrier, and one site of the well with one boson, is $\lambda$. This energy can be compensated if the barrier particle 
jumps to the single occupied site, so the energy of the two sites is $U$, and the transition happens for $\lambda \approx U$. Note that other 
possibilities are more expensive, which is corroborated by the phase diagram of the $AB_{2}$ chain shown in Fig. \ref{diag3}, where the gap tends to 
saturate for larger values of $\lambda$, banning quantum transitions in this region. A quantum phase transition around $U$ for non-integer densities 
is a newly discovered fact, due to the $AB_{2}$ chain.\par 
In Figures \ref{corr3}(c) and \ref{corr3}(d), we see that for the density $\rho=5/3$ the arrangement of the particles passes from one particle in the 
barrier and four in the wells to five particles occupying two sites in the wells, and the barrier site is empty when  $\lambda/t$ varies from 
$10$ to $24$, respectively. This situation is similar to the $AB$ chain case for the density $\rho=3/2$; therefore the quantum phase transition takes 
place around $2U$.\par 
Finally, for $\lambda/t=15$ and $\rho=2$, the configuration for the first CDW phase is shown in Fig. \ref{corr3}(e), with five particles in the wells 
and one in the barriers, but at $\lambda/t=24$, Fig.  \ref{corr3}(f) shows six particles located at two sites with the barrier site empty. The values 
of $\lambda$ indicate that the quantum transition happens around $2U$, a value for which there is no quantum transition in the $AB$ chain case. On the basis of  
Fig. \ref{corr3}(e), note that the smallest number of particles in a site of the wells is two, and the barrier is always occupied. The 
energy for two sites, the barrier, and one site of the well with two bosons, is $U+\lambda$, while $3U$ is the energy of the two sites after the boson at the 
barrier jumps to the site with two bosons. The fluctuations will increase when $\lambda \approx 2U$, and a superfluid region appears between the two 
CDW phases. The above discussion allows us to conclude that for a $AB_{N-1}$ chain with a integer density $\rho$, there are $\rho+1$ insulator phases 
separated by superfluid regions; $\rho$ insulator phases are CDW, and one is Mott insulator. The values of $\lambda$ around which the transitions 
occur will be determined by the values of $N$ and $\rho$.\par
We show  the phase diagram at the thermodynamic limit for bosons in a $AB_{2}$ chain in Fig. \ref{diag3}. For commensurate densities 
lower than one, the system exhibits two insulator phases separated by a superfluid one; the size of these CDW regions increases as a function of 
$\lambda$, but quickly saturates. We believe that there is a non-zero finite value of $\lambda$ for which the CDW phase appears for the densities $\rho=1/3$ and 
$\rho=2/3$. Note that these phases remain stable for all values of $\lambda$ considered in this paper; 
i. e., for these densities the arrangement of the particles is always the same regardless of the depth of the potential.\par
The phase diagram shows us that for non-integer densities larger than one, there are two CDW phases, similar to the $AB$ chain case, but for the 
$AB_{2}$ chain, we obtained that the transition regions depend on the density. However, the transitions happen around multiples of $U$. When 
$\lambda=0$, the ground state is superfluid for the densities $\rho=4/3$ and $\rho=5/3$; however, we expect a quantum phase transition to a CDW phase 
for larger values of $\lambda$.\par 
For integer densities, we observe that the ground state is of the Mott insulator type at $\lambda=0$ for any density $\rho$. When $\lambda$ increases, we 
see that CDW phases can appear, specifically $\rho$ CDW phases for a fixed global density $\rho$ (see Fig. \ref{diag3}).\par 
In the present paper, we consider bosons in an inhomogeneous lattice. Information about the system can be obtained using the tools of information theory, 
which have been shown to be useful for determining critical points of systems with a harmonic potential, showing that there is a one-to-one 
correspondence between the local von Neumann entropy and the charge fluctuations~\cite{silva}. Fran\c ca and Capelle~\cite{sou11} showed that 
the average of local von Neumann entropy is able to indicate the critical points where the phase changes happen in an inhomogeneous system, which 
is given by: 
\begin{equation}
\epsilon^{*}=\frac{1}{L}\sum_{i}\epsilon_{\nu N}(i), 
\label{von}
\end{equation}
\noindent where $\epsilon_{\nu N}(i)=-Tr \sigma_{i} \log_{2} \sigma_{i}$ is the local von Neumann entropy, 
and $\sigma_{i}=Tr _{B} \sigma$ is the density matrix of a single site located at $i$, where B represents 
the environment with $L-1$ sites, and $\sigma$ is the density matrix of the whole system. \par
\begin{figure}[t]
\resizebox{0.45\textwidth}{!}{%
  \includegraphics{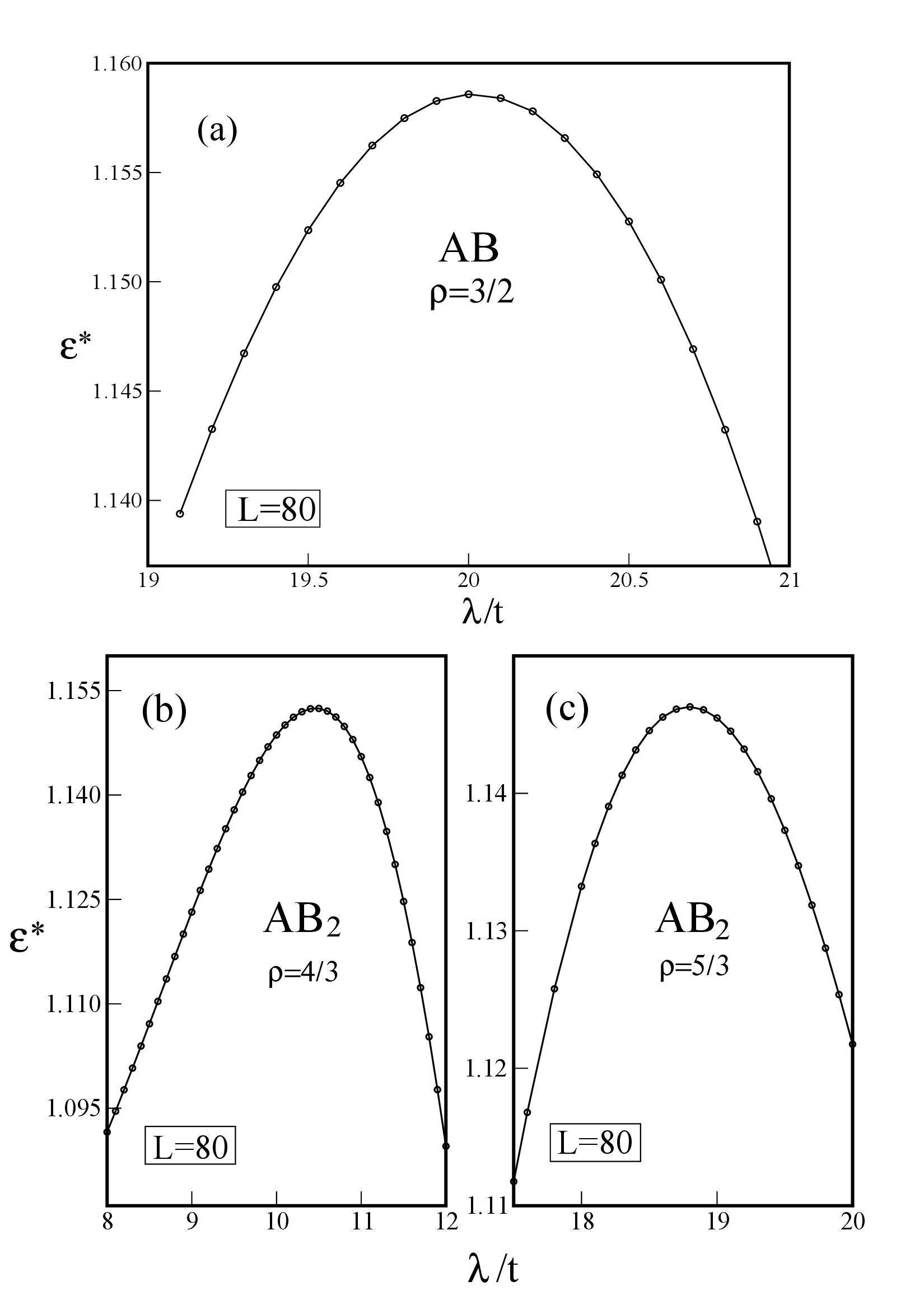}
}
\caption{Average entropy versus $\lambda$ for a $AB$ chain with density $\rho=3/2$ (a),
$AB_{2}$ chain with $\rho=4/3$ (b), and $AB_{2}$ chain with $\rho=5/3$ (c).}
\label{entropy}       
\end{figure}
In Fig. \ref{entropy}, we show the average von Neumann entropy versus $\lambda/t$. In (a) we consider the $AB$ chain, and the $AB_{2}$ chain in (b) and 
(c). In all cases, we consider densities between 1 and 2, and we observe the same behavior: the von Neumann entropy increases, reaching a maximum 
value, and then decreases. We note that it is impossible to determine the critical points that delimit the superfluid regions; however we can obtain 
information. In all cases, for values of $\lambda/t$ on the left side, the system has a characteristic distribution of particles, for instance 
$\{2,1,2,1,2,1.....\}$ for $\rho=3/2$ on the $AB$ chain. The average von Neumann entropy has a finite value associated with the number of degrees of 
freedom. When $\lambda/t$ increases, fluctuations also increase, and we expect that the average von Neumann entropy will grow. This tendency 
continues until the system reaches its maximum number of degrees of freedom, which would be associated with a coherent behavior of the particles, 
i. e. with a superfluid state. Note that the positions of the maximum values of the average von Neumann entropy agree with the middle point of the 
superfluid regions in the phase diagram for each density. Further increase in $\lambda/t$ lets the particles localize, and a different arrangement of 
the particles begins. Therefore, the number of degrees of freedom decreases and the average von Neumann entropy is smaller. An important fact is that 
the slopes before and after the maximum are always different, which indicates that the insulator regions surrounding the superfluid phase are 
different. For instance, for $\rho=3/2$, after the superfluid region the distribution of particles is $\{3,0,3,0,3,0.....\}$. This behavior was also 
obtained for integer densities.\par
\begin{figure}[t]
\resizebox{0.45\textwidth}{!}{%
  \includegraphics{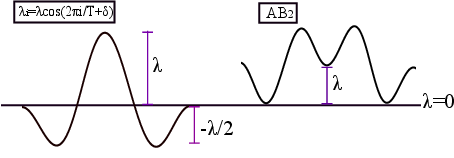}
}
\caption{Harmonic potential used by Rosseau \textit{et al.}~\cite{Rousseau} and Li \textit{et al.}~\cite{Li2015}, and the $AB_2$ potential 
considered here. }
\label{HvsAB2}       
\end{figure}
During the final editing of this paper, we became aware of a Li \textit{et al.} manuscript~\cite{Li2015}, who studied a boson system that undergo 
an external potential $\lambda_i=\lambda \cos(2(i+1)\pi/3)$ using quantum Monte Carlo simulations. They found a phase diagram which is 
qualitatively similar to the $AB_{2}$ phase diagram reported here, i. e. they show insulator phases for commensurate densities, multiples of 
$1/3$, and observe that for $\rho=1$ and $\rho=4/3$ a superfluid region separates two insulator ones. They analyzed the topological properties of 
the insulators and found topological nontrivial and trivial states; specifically, the Mott insulators are trivial.\par
The phase diagrams shown in the two papers are complementary, and we will use the Berry phase to determine the boundaries of the intermediate 
superfluid phase for densities $\rho\geq 1$, which results will be reported soon~\cite{CruzBarry}.\par
\begin{figure}[t]
\resizebox{0.45\textwidth}{!}{%
  \includegraphics{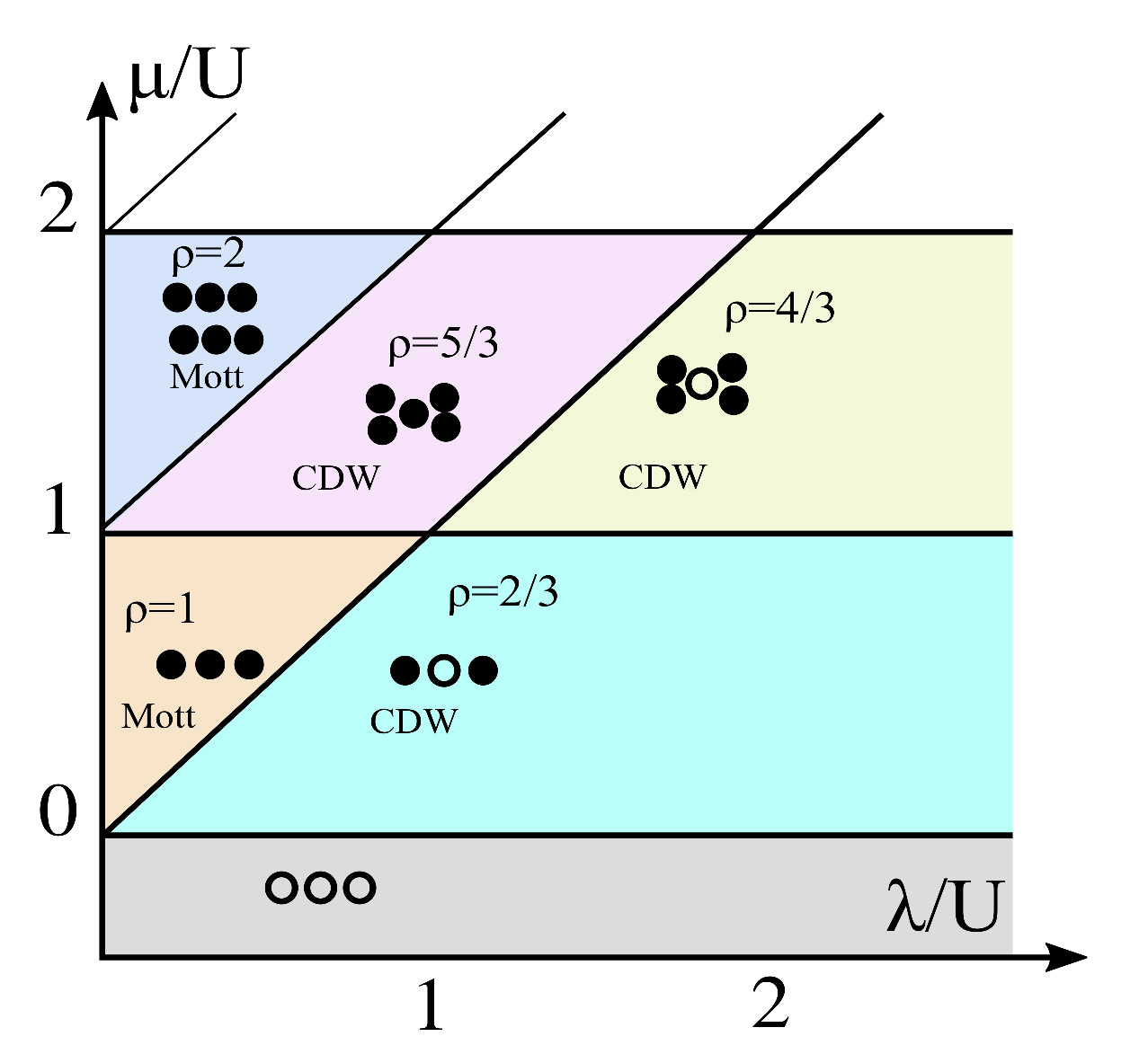}
}
\caption{(Color online) The phase diagram of bosons on an $AB_2$ chain in the atomic limit. Here $\circ$ represents a empty site and 
$\bullet$ a boson. 
}
\label{Latomico}       
\end{figure}
It is important to observe that in the phase diagrams found by Rosseau \textit{et al.}~\cite{Rousseau} and Li \textit{et al.}~\cite{Li2015}, who used 
the external potential $\lambda_i=\lambda \cos(2i\pi/T + \delta)$ ($T$ is the period and $\delta$ a phase), the 
phase boundaries for the lower insulating phases have a negative slope, whereas in our phase diagrams, the slopes are positive and the borders 
tend to be constant as $\lambda$ increases. To understand what is happening, we must first note that the potentials are different (see 
Fig. \ref{HvsAB2} ) and then consider the atomic limit $t=0$ of Hamiltonian (\ref{H1}). The ground state energy at site $i$, with $n_i$ bosons, is 
$E(n_i)=0.5Un_i(n_i -1) +\lambda_in_i-\mu n_i$. The energy difference between the energy with $n_i+1$ and $n_i$ bosons is 
$\Delta E=Un_i+\lambda_i-\mu$. When the result is zero, the border between insulating phases with different density is determined. If we consider the unit 
cell of the $AB_2$ chain (Fig. \ref{super}(b)) to be $BAB$, the occupation at site $A$ is determined by  $\mu/U=\lambda/U+n_i$. This relation 
generates parallel lines in the $(\mu/U,\lambda/U)$ plane that separate regions with different numbers of particles. However at site B, 
$\lambda=0$, and the occupation is given by $\mu/U=n_i$, which generates horizontal lines for each filling of site B. These lines are independent of 
$\lambda$, a fact that points out the main difference between our phase diagrams and the ones found by Rosseau \textit{et al.}~\cite{Rousseau} and 
Li \textit{et al.}~\cite{Li2015}. The phase diagram of bosons in an $AB_2$ chain at the atomic limit is shown in Fig. \ref{Latomico}. Here it can be 
seen that for larger values of $\lambda$ the borders between regions will be horizontal lines, in a manner similar to what we show in our DMRG 
calculations, which confirm the continuity of the phase diagram from the atomic limit to the delocalized case.\par 

\section{Conclusions}
\label{sec:sec6}
Using the density matrix renormalization group method, we determined the chemical potential  of $AB_{N-1}$ chains at the thermodynamic limit and 
found the phase diagram for the $N=2$ and $3$ cases. For a small energy difference between the $A$ and $B$ sites ($\lambda$), we observe that insulator 
regions with peculiar charge distribution (CDW) appear for densities $\rho=\alpha/N$ ($\alpha$ an integer), except at integer densities, for which 
the Mott insulator phase still appears. The size of the new CDW phases for densities less than one grows with $\lambda$, but then it stabilizes, and these 
phases remain in the phase diagram. On the other hand, for densities $\rho\geq 1$, we always found superfluid regions that separate two insulator 
phases, a result that was confirmed using the von Neumann entropy. Regardless of the value of $N$, we found that for integer densities $\rho$, there 
are $\rho +1$ insulator phases, these being $\rho$ CDW phases and one Mott insulator phase with $\rho$ particles per site.\par  
For $AB_{N-1}$ chains, we observed that there are two CDW phases for any 
non-integer densities larger than one. This result can be generalized by saying that for non-integer densities larger than one, there are $\rho +1$ CDW 
phases, where $\rho$ corresponds to the previous integer density. For instance, for commensurate densities larger than two, we expect three CDW 
phases.\par
The superfluid regions that separate the CDW phases occur around multiples of the local repulsion $U$, but the specific values and the size of these 
regions depend on the global density $\rho$ and $N$.\par 
 
\section*{Acknowledgments}
The authors are thankful for the support of DIB- Universidad Nacional de Colombia and COLCIENCIAS (grant No. FP44842-057-2015). Silva-Valencia and 
Franco are grateful for the hospitality of the ICTP, where part of this work was done. G. J. Cruz programmed the DMRG code and carried out the 
calculations. R. Franco contributed to the discussions. J. Silva-Valencia planned and designed the study. All contributed to writing the paper.

%
%

\end{document}